\def\bn{\bigskip\noindent}

\def\\{$\backslash$}
\def\hh       {{$^{h}$}}
\def\mm       {{$^{m}$}}
\def\ss       {{$^{s}$}}                  
\def\deg      {{\ifmmode^\circ\else$^\circ$\fi} } 
\def\arcm     {{\ifmmode {'\ }\else$'     $\fi} } 
\def\arcs     {{\ifmmode{''\ }\else$''    $\fi} } 

\def\cge      {{$_ >\atop{^\sim}$}}

\def\nref  {\noindent\parshape 2 0.0 truein 06.5 truein 0.4 truein 06.1 truein}

\documentstyle [12pt,aasms4]{article}

\begin{document}

\title{The Nature of Radio Emission from Distant Galaxies:
 The 1.4 GHz Observations}

\author{E. A. Richards}

\affil{National Radio Astronomy Observatory\footnotemark[1] and University of Virginia}
\footnotetext[1]{The National Radio Astronomy Observatory is a facility
of the National
Science Foundation operated under cooperative agreement by Associated
Universities, Inc.}
\authoraddr{
 520 Edgemont Road, Charlottesville, Virginia 22903 \newline
Electronic mail: er4n@virginia.edu}

\begin{abstract}

	We have conducted a deep radio survey with the
Very Large Array at 1.4 GHz
of a region containing the Hubble Deep Field. This
survey overlaps previous observations at 8.5 GHz
allowing us to investigate the radio spectral properties
of microjansky sources to flux densities greater than
40 $\mu$Jy at 1.4 GHz and greater than 8 $\mu$Jy at 8.5 GHz.
A total of 371 sources have been catalogued at
1.4 GHz as part of a complete sample within
20\arcmin ~of the HDF. The differential source
count for this region is only marginally sub-Euclidean
and is given by
$n(S) = (8.3 \pm 0.4) S^{-2.4 \pm 0.1}$
sr$^{-1}$Jy$^{-1}$. Above about 100 $\mu$Jy the radio
source count is systematically lower in the HDF as compared
to other fields. We conclude that there is clustering
in our radio sample on size scales of 1\arcmin - 40\arcmin .

	The 1.4 GHz selected sample shows that
the radio spectral indices are preferentially
steep ($\overline{\alpha } _{1.4} = 0.85 $) and the sources
are moderately extended with average angular
size $\theta$
= 1.8\arcsec . Optical identification with
disk-type systems at $z \sim $ 0.5-0.8
suggests that 
synchrotron emission, produced by supernovae
remnants, is powering the radio emission
in the majority of sources.
The 8.5 GHz sample contains primarily moderately flat
spectrum sources ($\overline{\alpha } _{8.5} = 0.35$),
with less than 15\% inverted. We argue that
we may be observing an increased 
fraction of optically thin bremsstrahlung
over synchrotron radiation in these distant
star-forming galaxies. 

\end{abstract}


\section{Introduction}

        Deep radio surveys show a surface density of radio
objects approaching 60 arcmin$^{-2}$ down to 1 $\mu$Jy (as
inferred from fluctuation analyses on the most sensitive
VLA images at 1.4 and 8.4 GHz),
or equal to the surface density of B = 27 magnitude galaxies
(Fomalont {\it et al.} 1991, Windhorst {\it et al.} 1993, 
Richards 1996, Richards {\it et al.} 1998, hereinafter
Paper I).
Despite their modest average luminosity of L$\simeq$ 10$^{22.5}$ 
W/Hz, the sheer number of microjansky sources implies that
they dominate the radio luminosity budget of the universe
at centimeter wavelengths.
However, surprisingly little is known about the physical
origin of these objects or their nature.

     Some generalizations from studies involving deep 5 and
8.5 GHz VLA fields imaged with HST and ground-based telescopes
are possible (Hammer {\it et al.} 1995, Windhorst {\it et al.} 1995,
Fomalont {\it et al.} 1997, Richards {\it et al.} 1998).
At least half of microjansky sources are associated with morphologically
peculiar, merging and/or interacting galaxies with evidence
for active star-formation (blue colors, infra-red excess, HII-like optical
spectra).  The remaining identifications are composed of low-luminosity FR Is,
ellipticals,
Seyferts, LINERs, and luminous star-forming field spirals. Thus a variety of 
physical mechanisms may be driving the observed evolution
among microjansky radio
sources, including non-thermal radiation from AGN activity, synchrotron emission
associated with diffuse supernova remnants, and thermal 
emission from HII regions.

        Another clue is available from the observed distribution 
of radio spectral indices. While at millijansky levels, the average 
spectral index for radio sources is about $\alpha \sim$ 0.8 
(Donnelly {\it et al.} 1987), microjansky sources
selected at high frequencies ($\nu \geq$ 5 GHz) have a
surprisingly flat spectra of $\alpha$ = 0.3 $\pm$ 0.2
 (Fomalont {\it et al.}
1991, Windhorst {\it et al.} 1993). Several explanations
for this observed flattening compared with sources selected
at higher flux density levels are possible,
including, free-free absorption, an increasing number of synchrotron
self-absorbed AGN among the microjansky population, and/or
a rising component of thermal radiation from active star-formation.


	The purpose of this 
study is to enlarge the faint radio 
sample and investigate the radio spectral  
and morphological properties for a statistically 
significant sample of microjansky sources.
In addition, examination of the optical
properties of the identifications may shed insight
into the nature of the radio emission in these
sources. To further this goal we have observed
a region of the sky at 1.4 GHz and 8.5 GHz centered on the
Hubble Deep Field, where excellent wide field optical
data are available. 
In Paper I (Richards et al. 1998) we
presented the optical identifications for a
complete sample of twenty-nine 8.5 GHz selected radio 
sources in our radio survey of the Hubble
Deep Field. The principal conclusion is that 
the majority ($\sim$70\% ) of the identifications
are with relatively bright (R$\sim$ 22 mag.) 
disk systems, many at moderate redshifts
($z$ = 0.4-1). In this paper (Paper II) we
present the 1.4 GHz observations. In section two
we describe our observations and data reduction
techniques. Section three presents
our source list, while in section four we
calculate the spectral index distribution.
In \S 5  we discuss the spatial clustering 
of sources in our catalog.
Finally, in \S 6 we summarize our
findings and give conclusions.

\section{The 1.4 GHz Radio Observations and Data Reduction}

	In November 1996, we observed a field centered
on the Hubble Deep Field ($\alpha$ = 12\hh 36\mm 49.4\ss \hskip5pt
and $\delta$ = 62\deg12\arcmin 58.00\arcsec \hskip5pt (J2000))
for a total of 50 hours at 20 cm in the A-configuration
of the VLA. In order
to minimize chromatic aberration, we observed in a pseudo-continuum,
spectral line mode with 7 $\times$ 3.125 MHz channels centered on
intermediate frequencies 1365 MHz and 1435 MHz, frequency windows
known to be relatively free of radio frequency interference.
Each frequency channel was composed of two independent
circular polarizations. 
All knowledge of
linear-polarized intensities was lost due to the limited
number of VLA  correlator channels. Visibility data were
recorded every 3.3 sec from the correlator.

	We monitored the point source 1313+675 (S$_{1.4}$ = 
2.40 Jy), located 6.5 degrees from the HDF, every 40 
minutes to provide amplitude, phase, and bandpass calibration.
Daily observations of 3C286 with assumed flux densities of
14.91 Jy at 1385 MHz and 14.62 Jy at 1435 MHz provided the 
absolute flux density scale.

	The calibrator observations allowed us to 
identify baselines with systematic phase or amplitude
errors. A few baselines were found to have recurrent
amplitude and/or phase closure errors greater than 
five percent and/or five degrees. All data associated
with these suspicious baselines were discarded.
To remove bursts of radio frequency interference,
we excised all visibilities with flux densities greater 
than about 10 $\sigma$ above the expected rms value of
0.08 Jy. This amounted to about 2\% of the data.

	After time averaging the $(u,v)$ data from 3.3 to 13 sec, 
we made preliminary 10\arcsec ~resolution maps 
which cover the field out to the first sidelobe of the 
primary beam about 0.8\deg ~from the phase center.
We searched these images for bright, confusing sources 
whose sidelobes might contaminate the noise properties
of the inner portion of field. All sources above 0.5
mJy were catalogued.
	
	We then imaged and heavily CLEANed these 
sources using the full unweighted $(u,v)$ data 
set. Because the primary beam response 
changes significantly over our 44 MHz bandpass (about 3\%),
it was necessary to deconvolve each of the confusing 
sources using each 3.125 MHz channel. In addition,
the confusing sources were independently deconvolved
in each of the circular polarizations
(right and left)
to account for the 'beam squint' of the VLA
antenna.
Their CLEAN components were Fourier transformed 
and then subtracted from the visibility data. Using 
these ''strong source subtracted data sets'', we then 
imaged the inner few arcmin of the field.
With this procedure the rms noise was found to be about
50\%
higher than expected from receiver noise alone.
In particular, a few sidelobes from
particularly strong sources (S $>$ 10 mJy) located near
the half power point and first null of the primary beam
were still apparent.

	By examining the images made from 30 minute
segments of the data, we isolated a few time intervals
where the visibility data appeared to be corrupted, 
possibily due to low level interference. For any 
of these 30 minute snap-shot images with a 
a rms noise greater than 50\% of the mean value,
the corresponding visibility data were deleted
from the final analysis.
These data amounted to about seven hours of time;
thus, in all, about 42 hours of
of high quality data were used to construct the final
images. The final combined data set has a rms noise
of 7.5 $\mu$Jy, whereas we expected a noise
closer to 5 $\mu$Jy.

\subsection{Construction of the 1.4 GHz Images}

	Our goal is to map the 40\arcmin ~field
of the 1.4 GHz observations out to the 20\% response
of the VLA antennas. There are two complications which
make this a difficult task. 

	First is the sky
curvature.
While in practice the VLA
is often treated as a two dimensional array, in
reality the instrumental response to radio emission
from the sky is a three dimensional
complex function. For observations of short duration, 
small fields of view, or low resolution, most
sources can be adequately deconvolved without reference 
to the so called 3-D effect. However, the isoplanic
assumption fails for sources located 
further than $\theta \simeq 1/n_{syn}$ radians from
the phase tracking center,  where $n_{syn}$
is the phase center distance in synthesized beam widths.
For the A-array at 1.4 GHz this corresponds to a 
patch size of about 18 arcmin across.

	Thus we chose to approximate the one degree 
primary beam of our observations by using a number
of independent and equidistant facet images. Each
facet is constructed from the Fourier transform of
the data which has been phase shifted to its tangent
point on the celestial sphere. In all, 16 facets
were used. This technique is
known as polyhedral imaging and is discussed at length
by Cornwell \& Perley (1992).

	One further complication comes from the practical
limitation of the extensive computations needed to reduce
these these wide
field observations. Our entire calibrated and edited
data set consists of over $10^7$ discrete complex 
visibilities (after 13 sec time averaging) which must be 
directly Fourier transformed to compute the sky brightness
distribution over approximately the same number of pixels.
Because we are interested in radio emission over the
entire primary beam, we must properly deconvolve sources
within the entire field of view. Although only sources fainter
than 0.5 mJy remain in the visibility data after
the subtraction of the confusing sources, their
collective sidelobes can limit the dynamic 
range of the final images. 

	The most accurate method for deconvolution of the
faint sources is to CLEAN each of the 16 images in
parallel, subtracting each source's sidelobe contribution
from all the image facets simultaneously.
In this manner, one would recover the sky brightness
function free from sidelobe contamination over the
entire primary beam. However, in practice, this
is a prohibitive computational task. Therefore we opted
to CLEAN each of the 16 facets in series using
the much more efficient  Clark-Hogbom algorithm 
(Clark 1980). The price paid is that sidelobes from 
the multitude of sources less than 0.5 mJy are only 
properly removed locally, within the individual image facet
which contains each source. 
	
	In order to examine what effect this might have
on the rms noise near the center of the image, we
performed a simple test. First we made an image 
of the inner seven arcmin of the field center, heavily CLEANed
using the Clark-Hogbom method. Then we made a similar image
twice the size of the former, again heavily CLEANed. 
Comparison of the two images, one relatively free of sidelobe
contamination from the far field sources with S $\leq$ 0.5 mJy,
the other not, yields a first order approximation
of the unCLEANed sidelobes left in the center of the
field due to our deconvolution method outlined above.
About 0.8 $\mu$Jy of flux density remains per independent
beam, and thus increases the noise about 10\% over the
thermal noise.

	We produced the final image by CLEANing each of the
 sixteen 2048 $\times$ 2048 pixel fields (0.4$\arcsec$ pixels) with
10,000 iterations and a 10\% loop gain. This produced
an image with a natural resolution of  $2.0\arcsec \times 
1.8\arcsec $ and P. A. = $-86\deg.$ The rms noise
near the center of the field was 7.5 $\mu$Jy,
approximately 30\% higher than the value expected
from thermal noise alone based on our system temperature
(37 K on average),
integration time, and bandwidth.
The noise in each of the 16 images was found to 
be fairly uniform, although it increased to as much
as 10 $\mu$Jy in regions near the brightest
millijansky sources. 

	Residual ringlike structure from 
unCLEANable sidelobes around these strong sources
was evident, suggesting that our images are dynamic
range limited at about 5000:1. This is typical
of blank field deep imaging where the data cannot
be effectively self-calibrated.
We clipped out regions
immediately around these sources, as well as a
larger region around one particularly strong source
(S = 35 mJy) located at the half-power point
of the primary beam. We attribute the dynamic
range limitation in this image to pointing 
fluctuations, which are typically 
15\arcsec ~rms for the VLA. This effect causes
amplitude fluctuations in the apparent brightness
of the radio sources,
which are of order 1\% at the half power
point, and hence induces unCLEANable sidelobes
into the image. 

        Examination of the stronger deconvolved
sources in the field with $S _p$ \cge 5 mJy showed evidence
of a radial sidelobe oriented towards the phase
tracking center, or center of our images. These
sidelobes appeared to be symmetric about these
stronger millijansky sources in the field and with first
order amplitude of about 10\%. Their width was
approximately that of the delay beam.

        These sidelobes were also apparent during
our test observations of 1400+621, and
only appear around the off axis observations. Other
observations taken at the VLA have not shown similar
artifacts and our suspicion is that the problem
was an online system recording error with the 3.3 sec
visibility data that we chose to use. Examination of
VLA data of the same test field taken in an
identical observing mode, except with 10 sec visibility
integration,
showed no sign of the radial sidelobes.

        The cause of these artifacts is unknown.
They do not appear to be present around the near field,
weaker sources (less than about 1 mJy). As a
test of whether these sidelobes might adversely effect
our measurements of the flux densities of the stronger
millijansky sources, we examined the off-axis flux
density measurements of 1400+621 in images with and
without the presence of the radial sidelobes.
The measurements agree to within the standard flux
density errors. It is more difficult to assess
if these radial sidelobes adversely affect the general rms noise
in the final 1.4 GHz images.

\section{The Complete Source List}

	In order to minimize the introduction of
spurious radio sources into our sample, we 
examined the distribution of the $negative$ pixels
in the images to estimate its completeness limit.
 In general most regions appeared
to reflect well behaved Gaussian noise with the 
most negative peak being about five times the local
rms noise. However, one region with a local rms
of 7.5 $\mu$Jy contained an unresolved negative feature
with $S_p$ = -63 $\mu$Jy. This feature appeared
to be quite isolated and located several arcmin 
from any of the brighter millijansky confusing sources.
No other strange artifacts (e.g., rings, streaks,
residual sidelobes) were apparent in the vicinity of
this negative 'source'; hence its presence remains
enigmatic. The next most negative pixel value
of -40 $\mu$Jy is located near a strong confusing
source. We therefore adopt 40 $\mu$Jy to be the formal 
completeness limit over our entire one degree field.
Even if there are positive counterparts to the
 -63 $\mu$Jy 'source', there should be no more than 
a few if the negative and positive pixel amplitude 
distribution are fairly symmetric about zero. However,
we note that the probability of finding a --9$\sigma$
source within our field is much less than one percent,
and demonstrates that the noise properties of this
image are not entirely Gaussian.

	Next we searched our images for all 
pixels with S $\geq$ 40 $\mu$Jy and fit 
these sources with elliptical Gaussians to determine
their peak flux densities and positions using the
automated AIPS task SAD. In all 314 sources 
were found within 20 arcmin of the center of the 
image (20\% power contour). 

\subsection{Determination of Discrete Source Angular Sizes and Flux Densities}
	
	Gaussian fitting routines such as SAD are subject to
noise dependent biases which cause significant overestimation
of source sizes and flux densities (Windhorst et al. 1984, Condon 1997). In order to estimate 
the effects of population and noise bias in our images, 
we performed a series of Monte Carlo simulations. 
Our basic technique was to inject a number of point
sources (100) of known flux density into the CLEANed sky images
of this field. Then these sources were 
recovered from the images using
SAD and their measured peak and integrated flux densities compared
to the input model. We used the ratio $S_p /S_i$ to determine
if the simulated source was significantly broadened by
population and/or noise bias. 

	Because our goal is to set a resolution criteria
in the presence of these fitting biases, we performed
this simulation as a function of flux density, or
alternatively, signal to noise. By examining the 
distribution of $S_p /S_i$ at different flux densities, we obtained some idea
about at what level we could confidently believe that a given source
was resolved in our real sky images. 
For a source of intrinsic dimension $\theta _{maj} \times \theta _{min}$
resolved by a beam of extent $a _{maj} \times a _{min}$,
the integrated and peak flux densities are related by
$S_i = S_p (1 + \frac{\theta _{maj} \times \theta _{min}}{
a _{maj} \times a _{min}})$, thus allowing us to relate
the ratio $S_p /S_i$ directly to a source angular size. 
By examining the distribution of modeled $S_p /S_i$
to input $S_p /S_i$, we were able to adopt a 
95\% confidence in the resolution criteria as a function of signal to noise.
The resulting best fit data yields
$S_p /S_i$ (95\%) $< 1 - \frac{2.3}{\sigma _{snr}}$
where $\sigma _{snr}$ is the signal to noise ratio of the
radio source. Thus at the detection limit of our 
images (40 $\mu$Jy), a source must have a $S_p /S_i$
value less than 0.57 to satisfy our resolution criterion,
or an angular size greater than about 2.7\arcsec .
It follows that we can resolve only those sources
stronger than 70 $\mu$Jy
integrated flux in our complete sample.
	
	Next, we used these same simulations with
the sources of angular size that we could have just 
detected at each signal to noise ratio (e.g.,
2.7\arcsec ~at $S_p$ = 40 $\mu$Jy). Comparison
of the input to recovered peak and integrated flux
densities, and angular sizes yields an estimate 
of the bias induced by Gaussian fitting techniques
in the presence of population and noise bias. We corrected
our real sky source parameters to account for these
biases. In general the peak flux densities of the fitting routines
were in good agreement with the models, while the integrated
flux densities and angular sizes from the Gaussian fits
 were overestimated by up to
a factor of 1.5.

	As a check on the integrated flux densities for 
apparently resolved sources
in our complete sample obtained from the 
Gaussian fitting algorithms, we examined the distribution of
peak flux densities as measured in various resolution 
images. Images were constructed at 1\arcsec , 3.5\arcsec , 
and 6\arcsec ~ resolution in addition to our nominal
2\arcsec ~naturally weighted image. Table 1 gives the 
parameters of each image. For those sources which satisfied
our initial resolution criterion, we checked that the peak
flux density of the source increased with decreasing resolution
consistent with the fitted angular size. Those sources
which did not were considered unresolved
and their adopted peak and integrated flux densities were 
measured on the 
the 2\arcsec ~image. Figure 1 shows a greyscale of
the inner 10\arcmin ~of the 1.4 GHz image. In
Figure 2, we show the greyscale of the 8.5 GHz mosaic
image (see \S 6).

	We also searched the lower resolution 3.5\arcsec
~and 6\arcsec ~images for resolved sources
not detected in the untapered 2\arcsec ~image above
our completeness limit of $S_p$ = 40 $\mu$Jy. This search
yielded 57 additional sources in the tapered
images above completeness limits of 50 $\mu$Jy
at 3.5\arcsec ~resolution and
75 $\mu$Jy at 6\arcsec ~resolution.
Because the angular sizes are uncertain in these 
low signal to noise ratio detections
(and in many cases may be instrumentally broadened), the
adopted peak and integrated flux densities were set equal
to the peak pixel values in the tapered images.
These additional
sources were added to the complete source list which
in total contains 371 sources. The complete sample
of all radio sources detected at 1.4 GHz within 20\arcmin
~ of the phase center is presented in Table 2. 

A description of Table 2 is as follows.  All uncertainties are given
at the one sigma level.

{\em Column} (1) --- The Right Ascension in J2000
                     coordinates with one sigma uncertainty.

{\em Column} (2) --- The Declination in J2000
                     coordinates with one sigma uncertainty.

{\em Column} (3) --- The {\em deconvolved} (FWHM) major axis
                     of the best Gaussian fit to the source,
                     $\Theta$, is given in arcsec.

{\em Column} (4) --- The signal to noise ratio of the detection
                     calculated from  $S_p /\sigma $
                     where $S_p$ is the peak flux density as
                     measured in either the 2\arcsec , 3.5\arcsec ,
                     or 6\arcsec ~image, and $\sigma$
                     equals the rms noise in that image.

{\em Column} (5) --- The integrated sky flux density (S$_{1.4}$)
                       after correction for the instrumental gain,
  (see \S 3.2),

                     with corresponding one sigma errors.

\subsection{Instrumental Corrections}

        We must also correct the derived source
parameters for various instrumental effects.
In order to measure the off axis response of the
VLA to a point source, we observed the 4.2 Jy point
source 1400+621 in
each of the cardinal directions at positions of 5, 10,
15, 20 and 25 arcmin from the nominal phase
reference center.

        There are four principal effects which
reduce point source response as a function of
radial distance from the phase center. 
In their degree of increasing importance
they are
1) 3-D smearing, 2) finite time
visibility sampling (time delay smearing), 3) chromatic aberration
(bandwdith smearing),
and 4) primary beam attenuation.

1. We have approximated the curvature
of the celestial sphere with a number of 
two-dimensional facets. However, smearing may still be
present at a reduced level near the edges of each 
of our individual facet images. The effect of a 
small amount of 
3-D smearing are such that the integral
flux density of a given source is preserved while its
peak signal is reduced by an amount dependent
on distance from the tangent point on the celestial sphere.
 In order to 
determine the amplitude of 3-D smearing in our data,
we performed a series of 
simulations inserting point sources at a 
variety of distances from the phase tracking center
into visibility data with our exact $(u,v)$ coverage.
These data were then imaged in the same manner 
as the true sky images.
In this manner we determined the amount of point 
source degradation as a function of distance from 
the celestial sphere tangent point or image facet 
center. At the maximum possible distance of a 
source from a tangent point ($\sim$ 600\arcsec )
the peak degradation is less than 20\%. All 
values of the peak flux density ($S_p$) were corrected
for this effect according to our empirically fit
polynomial.

2. The calibrated $(u,v)$ data set used to construct 
our images consisted of 13 sec averaged visibility data.
Because the actual radio sources rotate in the sky during
this sampling time, their flux density is smeared in the 
image plane. The analytic calculation of this
smearing is complicated by the aspects of the observing 
geometry. However, at the North Celestial Pole the 
effect reduces to a tangential smearing in the image
plane and its amplitude is given by $S_p/S_t$ = 1 --2.06
$\times 10^{-9}$ $\theta$/$\theta _{syn}$ where $S_t$
is the integrated flux density of a source located 
an angular distance $\theta$ from the phase tracking center and
$\theta _{syn}$ is the size of the synthesized beam.
We used this 
approximation to correct the peak flux densities
as measured in our images.

3.  Although we planned our observations of the HDF
field to minimize the effects of chromatic aberration, 
for far-field sources this effect can still be 
important. It is especially crucial to understand
the effects of smearing on the completeness level
of the images. 
We measured
the off-axis response of 1400+621 due to chromatic
aberration by examining the ratio of the peak
to integrated flux density, $S_p /S_i $, as a
function of distance from the image phase center.
Bandwidth smearing is governed by the
observing frequency, 
width and shape of the bandpass and the $(u,v)$ 
coverage, all of which define the synthesized beam.
In theory, these are known functions and the
final beam response can be calculated analytically
as a function of position in the image. However, in
practice such uncertainties as non-uniform $(u,v)$ 
coverage, central intermediate frequency offsets,
and imperfect bandpass filters
make this impossible. We chose instead to fit a function
to the empirical data of the form 
$S_p /S_i $ = (1+(r/k)$^2$)$^{-0.5}$ where r is the 
distance from the phase center and k is a constant which
absorbs the uncertainties discussed above. Our least 
squares fit to the empirical data is shown in
Figure 3 where k = 16.19 arcmin. This is within 3\% of the
theoretical value assuming $\delta \nu /\nu$ = 3.125/1400.5
and a $\theta _{beam}
$ = 2.1\arcsec . For purposes of 
correcting $S_p$ in our images, we use the above equation scaled by 
the appropriate beam size. The integrated flux density,
$S_i$ is preserved and needs
no correction for smearing.

4. The primary beam attenuation at 1.40 GHz has
been measured to about 1\% (Condon 1997) to
the first sidelobe (-10 dB). 
Each of the peak and integrated flux densities in the
source list were corrected for the primary beam attenuation.
The uncertainty in the correction is the standard rms pointing
error of the VLA elements (about 15\arcsec) multiplied
by the differential log of the primary beam response.

	In Figure 3 we plot the point source response
of a source as a function of distance from the phase
center. We define the instrumental gain factor as the
product of bandwidth smearing, time averaging smearing,
and the primary beam correction.


\subsection{Positional Accuracy}


	The assumed position of our phase
calibrator 1400+621 is
$\alpha$ = 14\hh 00\mm 28.6526\ss
\hskip5pt and $\delta$~=~62$^\circ$15\arcmin 38.526\arcsec \hskip5pt
(J2000). We observed systematic, monotonically 
increasing shifts in both RA and DEC of comparable amplitude
as a function of $r$. It is a small effect and the
difference between the actual radial distance from 
the field center and that measured in the image plane
is $\delta$r = -0.042 arcsec at 25 arcmin distance from
the field center. 

	This small systematic term can be explained by
the so called annual aberration effect (Fomalont et al. 1992).
The predicted scale
contraction from annular aberration in the direction
of the HDF during the Julian epoch 1996.6 is
0.9999783. The good agreement
between the observed and predicted scaling factor 
for 1400+621 yields confidence that our images are 
free from significant distortions due to asymmetric
bandpasses or IF offsets. 

	We also corrected all the radio source
positions in our catalog for position
offsets induced by the 3-D effect as discussed
in \S 3.2. By phase shifting the image tangent
points to the location of individual
sources, their true angular positions on 
the sky were measured.
The difference between the apparent
position as measured on the nominal
sky maps and the corrected positions
is typically only 0.2\arcsec ~
in the north-south direction.

	After correction for annular aberration
and the 3-D term,
the relative positional accuracy for
sources across the field in the limit of infinite
signal to noise should approach about 0.02\arcsec.
The absolute positional accuracy depends on the
translation of our phase calibrator position to
that of the HDF and is about 0.02\arcsec .
Thus we estimate our radio catalog to be
within 0.03\arcsec ~of the J2000/FK5
coordinate grid. The single coordinate
rms position errors as  given in Table 2 
are defined as $\sigma = \sqrt{(1.8\arcsec /2\sigma 
_{snr})^2 + (0.03\arcsec)^2}$.


\section{Survey Completeness and Source Counts}

    Because the completeness of the radio source
sample is defined in terms of peak image flux, $S_p$,
corrections must be made for the the instrumental response
and biases inherent in our detection algorithm.
Although we have corrected the individual sources in
Table 2 for these effects, we now calculate the
fraction of sky sources which remain undetected
in our survey due to the finite angular size of the
sources.

\subsection{Angular Size Distribution}

	Previous high resolution studies of the 
microjansky radio population suggested that the
median angular size ($\theta _{med}$) for submillijansky
radio sources is approximately 2\arcsec ~and
almost independent of flux density between
80 - 1000 $\mu$Jy (Windhorst et al. 1993, Fomalont
et al. 1991, Oort 1988). The 
resolution of our present microjansky survey
is thus well suited to study the angular size
distribution for a statistically large and
well defined sample of microjansky radio 
sources. 

	Of the 151 radio sources in our complete
sample with  70 $\mu$Jy $ <  S_i < 1000 ~\mu$Jy 
for which we have angular size information,
only 77 (50\%) are resolved
with our typical 2\arcsec ~ resolution
limit. We divided these sources into
two flux density bins, containing approximately 
equal numbers of resolved sources.
Considering only the number of radio sources
with angular size greater than 2.7\arcsec ~
(the angular size detection limit of the weakest
radio sources in our sample as defined in \S 3.1),
we find twice as many resolved sources with 
$S_i >$ 250 $\mu$Jy (52\% ) as opposed to those
with $S_i \leq $ 250 $\mu$Jy (25\% ).
This suggests that $\theta _{med}$ may be
a {\em decreasing} function of flux density.
Figure 4 shows our measurements of angular
size as a function of source intensity.
In order to estimate the 
mean angular size of this sample, we applied
the survival analysis techniques of Feigelson
\& Nelson (1985) using the statistical 
package ASURV (Rev. 1.2; Isobe \& Feigelson 1992).
This technique incorporates upper limits
in the calculation of the mean of a distribution,
which is particularly important in our sample
which is dominated by non-detections (i.e. we
are measuring the tail of the angular size
distribution). The technique assumes a symmetric
Gaussian model, hence the mean and median are 
equal.
At $S_{1.4}$ = 370 $\mu$Jy, $\theta _{med}$ = 2.6\arcsec
~$\pm$ 0.4\arcsec , and at $S_i$ = 100 $\mu$Jy
, $\theta _{med}$ = 1.6\arcsec $\pm$ 0.3\arcsec .
The errors are based on the number of angular
size measurements (not limits).

	In Fig. 5 , we compare our determinations of
$\theta _{med}$ with previous measurements made
at 1.4 GHz. Because of the uncertain selection effects
inherent in the higher frequency deep radio surveys,
particularly their bias towards flat-spectrum, compact
AGN, we chose not to include these points. With the notable
exception of the discrepant Condon \& Coleman (1985) point, there is 
general agreement amongst the different data. 
Because the median angular size is known to 
change rather sharply below a few millijansky
at 1.4 GHz (presumably due to the emergence of
an increasing population of starburst galaxies among radio sources)
from $\sim$10\arcsec ~to 
a few arcsec, we suggest that the high flux density point
of Oort (1988) is too low, possibly due to
resolution biases in his A-array snapshots.

	We believe the decrease in angular size at lower
flux densities to be real. 
Thus for the purposes of modeling the median
angular size-flux density relationship,
we fit a function of the form:

$\theta = frac{3}{4} \times 0.175 S_{1.4} ^{0.5}+ frac{1}{4} \times 0.1$ arcsec

	This fit is also shown in Fig. 5. For completeness,
we also plot a straight line, with $\theta $ = 2.0\arcsec ~
and independent of flux density. 

\subsection{Completeness}

	In order to investigate the combined 
effects of noise, population, and resolution
bias on the completeness level of our
survey, we used a series of Monte Carlo simulations
similar to the ones described in \S 3.1.
In particular we want to determine how many sources
with $S_i \geq$ 40 $\mu$Jy, but with $S_p <$ 40 $\mu$Jy
we missed based on our peak flux density
detection limit. We randomly populated
our sky images with
100 sources of
finite angular size assuming an angular
size distribution as found in \S 4.1.
This simulation was repeated at a variety of 
flux densities from 40 to 1000 $\mu$Jy. In each
flux density interval the ratio of the number of sources
recovered from the images with $S_p \geq$ 40 $\mu$Jy
to the number originally injected in the model was
tabulated. This was taken to be the effective 
correction factor needed to account for the combined
effects of resolution, population and noise bias
in our images (although resolution bias is always
the dominant term).
At 80 $\mu$Jy which is the average flux density
source detected in our survey (weighted by $S^{-2.4}$)
and where the count will most accurately be determined,
this correction factor is 1.05. As resolution
bias is the dominant source of incompleteness
in our survey, we estimate that we have detected
approximately 95\% of the microjansky sources
in the HDF region to this flux density limit.
The principal uncertainty in the correction factor
is the uncertainity in the angular size distribution of the microjansky radio 
sources. If we had assumed a constant
 $\theta$ = 2.0\arcsec ~model, our corrections
would have increased by over a factor of two.

	From the complete source list of Table 2,
we then binned sources in flux density intervals
such that each bin had at least 50 radio sources
(except for the highest flux density bin).
The differential count was then calculated
based on the number of sources in each bin interval.
Bandwidth smearing, time averaging
smearing, and the primary beam response
decreases the effective area over which a source of given
$S_p$ can be detected. Thus when counting the number of
sources in each bin, care must be taken to weight
the contribution of each source to the count by the effective
area over which it could have been detected. This factor
can be calculated by solving for the image radius where
a source of amplitude $S_p$ would have just been missed by
our peak detection limit (Katgert 1973). Table 3 presents the
differential source counts for our complete flux 
limited sample. The counts from this survey are compared
to other microjansky surveys in Figure 6a and 6b
(Mitchell \& Condon 1985, Oort \& Windhorst 1985, Hopkins
{\em et al.} 1998) normalized to a Euclidean
geometry n/$n_o = n(s) /S^{-2.5}$. In general
the agreement is reasonable and in agreement with 
the errors and possible field to field variations.
The best fit to the source counts in this field 
in the range 40 - 1000 $\mu$Jy is
$n(S) = (8.25\pm 0.42) S^{-2.38 \pm 0.13}$ ster$^{-1}$ Jy$^{-1}$.

	The counts in the HDF appear systematically lower
than those of other fields above 100 $\mu$Jy. This effect could be
due either to 1) real field to field variations 
on the degree scale as a result of large-scale
clustering of radio sources,
or 2) survey incompleteness due to the finite angular
size of the radio sources. Without complementary, low
resolution observations, it is difficult to discriminate 
between these two possibilities.
We note that if the mean angular size 
does not decrease significantly below 100 $\mu$Jy, the 
radio sky will become forever naturally confused at the 
level of a few hundred nanojansky, perhaps
providing a natural limitation to the sensitivity 
of the next generation of centimeter radio telescopes 
(Windhorst et al. 1993).

\section{Spatial Clustering}

	In order to test for the presence of two dimensional
spatial clustering among the radio sources in the
Hubble Deep Field, we calculated the two-point
correlation function for the sources in Table 2.
First, we compiled a table of angular separations
by considering the separation
of each individual source with all other sources
in the catalog. These provide our $DD$
estimate (Peebles, 1980). Next, we generated 
random catalogs of sources according to the
source count of \S 4.2 , and distributed randomly
across a 40\arcmin ~ VLA primary beam. The
peak flux densities of these sources have been
attenuated by the instrumental corrections
discussed in \S 3.2. Angular pairs were calculated
for these catalogs and form the basis of 
our $RR$ measurement. We define
the correlation function of our catalog 
to be $w(\theta )$ = DD/RR -1 . The 
correlation function of our catalog 
on scales of 0\arcmin - 40\arcmin ~
is presented in Table 4.

 	We calculated the errors in our clustering
measurement by following the bootstrap
method of Ling et al. (1986). These agree
well with the Poissonian error estimate
of $\delta w(\theta )$ = 1 + $w(\theta)$)/N$_{DD}$
where N$_{DD}$ is the number of independent data
pairs in a given bin. We find evidence for an
excess of radio sources on scales of 
approximately 1-10\arcmin, while on scales
much larger than this there are
fewer radio sources in our catalog than
expected from a random distribution. Figure
7 shows the correlation function for
radio sources in the HDF, compared to 
the correlation function of a somewhat shallower 
1.4 GHz survey of Oort (1987) complete 
to 100 $\mu$Jy. The amplitudes are comparable
in the two separate surveys.

	Spatial clustering at higher flux density
levels and lower amplitudes has been reported by 
Cress et al. (1996) and Magliocchetti et al. (1998).
More recently, Hopkins et al. (1999) claim 
fluctuations in field to field source counts
at similar completeness levels to ours,
possibly indicating the presence of large-scale
radio source spatial variations. Thus it is
plausible that there are both fewer radio sources
in the HDF region than the average field, and that these sources
are clustered on arcmin scales amongst themselves.

\section{Radio Spectral Indices}

	The HDF has been observed previously
with the VLA at 8.5 GHz to a one sigma sensitivity 
of 1.8 $\mu$Jy (Paper I).
In June 1997 we observed the HDF region for
an additional 40 hours at 8.5 GHz.
We mosaiced an area defined by four separate pointings
offset 2.7\arcmin ~from the center of the HDF (the half-power
scale of the primary beam response)
in each of the cardinal directions for about 10 hours
duration each. The observing technique and data
reduction are discussed in Paper I.
The final combined 8.5 GHz images have
an effective resolution of 3.5\arcsec ~and
a completeness limit of 8 $\mu$Jy. 
The sensitivity of this mosaic to sky emission
is a sharp function of distance from the
nominal pointing center because the observations
were heavily weighted towards imaging the 
central HDF region.

	Because the size of the VLA primary beam
scales inversely with frequency, our sensitivity
at 8.5 GHz is limited to the inner 6.6 arcmin (HWHM)
of the 1.4 GHz field. This is the point where
the maximum beam attenuation at 8.5 GHz is
equal to 0.2 (while at 1.4 GHz the attenuation is
only 0.9). Within this region there are 109
sources contained in the 1.4 GHz complete
sample. We measured the 8.5 GHz flux density 
at the location of each of these sources.
When a source was not clearly detected ($S_p <
3 \sigma$ at 8.5 GHz), we calculated a conservative
upper limit to its 8.5 GHz flux density 
equal to three times the rms noise corrected by
the antenna gain. If a 1.4 GHz radio 
source had a peak flux value 3$\sigma < S_p 
<  5 \sigma$, its flux limit was taken as $S_p$
also corrected by the primary beam. This ensures
that our 1.4 GHz selected spectral index 
sample is complete and free from uncertain
weak source biases.
Based on this criteria, 30 sources from the
1.4 GHz sample had clear counterparts in the
8.5 GHz image.  Using the 1.4 GHz and 8.5 GHz
flux density values as measured in their
respective 3.5\arcsec ~convolved images
, we 
calculated individual spectral indices
using the convention $S_{\nu} \propto \nu ^{-\alpha}$.
In the following discussion, steep spectrum 
sources are defined as those with $\alpha \geq$ 0.50,
while flat spectrum as those with $\alpha < $ 0.50.

	 The 1.4 GHz to 8.5 GHz spectral index
distribution of the 1.4 GHz selected sample
is shown in Figure 8 (only those with meaningful
lower limits, $\alpha \geq$ 0.50 are shown
for clarity). Because of the large
number of spectral index limits for the 1.4 GHz
sample (79/109), we chose to only consider
those sources with $S_{1.4} > $ 100 $\mu$Jy
when calculating the mean of the
sample. The mean as calculated
from both the detections and lower limits (using ASURV)
for sources with $S_{1.4} > $ 100 $\mu$Jy
is $\overline{\alpha } _{1.4}$ = 0.85 $\pm$ 0.16.
For those sources with $S_{1.4} > $ 100 $\mu$Jy
the fraction of steep spectrum sources is
$\alpha$ = 0.62.
We also calculated the median of the spectral
index distribution for the entire 1.4 GHz sample.
 We did not consider 
spectral index lower limits which were weaker than
${\alpha } _{1.4} >$  0.3, to avoid a
bias in the median calculation. The median
of the 1.4 GHz selected sample is 0.63.

	We now consider those radio sources 
within the central 6.6\arcmin ~(HWHM) detected
on the basis of the 8.5 GHz data alone.
There are 29 sources in the complete sample
8.5 GHz sample of Paper I.
Eleven additional sources were detected in
the mosaiced regions with $S_p \geq 5 \sigma$.
These 40 sources comprise a complete sample of
radio sources detected at 8.5 GHz within 
6.6\arcmin ~of the HDF center. 
We measured 
the 1.4 GHz flux densities at their positions
in the 3.5\arcsec ~image. All but 10 of
these sources are contained in the complete 1.4 GHz
sample in Table 2. Upper limits to the 1.4 GHz
flux density were calculated as three times the
rms normalized by the antenna gain.

The spectral index
distribution is shown in Figure 9. The mean
spectral index for the 8.5 GHz selected sample
is $\overline{\alpha } _{8.5}$ = 0.35 $\pm$ 0.07 while the
median spectral index of the detections is 0.41.
The fraction of flat spectrum sources in the
8.5 GHz selected sample is 0.60.
Table 5
gives individual spectral indices or limits
(where meaningful limits are available)
for both the 1.4 and 8.5 GHz samples.

\subsection{The Nature of Flat Spectrum Sources: AGN vs. Starbursts}

        It has been noted by previous authors
that below a few milljansky, the 
median spectral index for high frequency selected
samples ($\nu \geq$ 5 GHz) flattens from a value of
0.7 to about 0.3-0.4 and then remains constant
for at least two decades in flux density 
(Windhorst et al. 1993 and references
therein). Our spectral index study confirms
this trend.

This raises the question of what physical
mechanism is responsible for the flattening of
the high frequency selected microjansky
population. One clue comes from the optical
identification of the sources. Of the 26
flat spectrum sources presented here, only
4 can be reliably associated with elliptical
galaxies, the majority (70\%) residing
in mergers, interacting disk systems, or
isolated field spirals (Paper I, Richards et al.
1998). Thus the flattening of
the spectral index distribution for the microjansky
population is unlikely to be due to radio evolution of
the elliptical population.

If the 8.5 GHz
sample preferentially selects out
self-absorbed AGN cores from the microjansky
population, then we might expect these sources
to have a smaller angular size on average
as compared to a 1.4 GHz selected sample.
The mean angular size for the 26 flat
($\alpha < 0.5$) spectrum
sources of Table 5, is $\theta $ = 1.7 $\pm$ 0.6,
as compared to $\theta $ = 1.8 $\pm$ 0.5, for
the 47 steep ($\alpha \geq 0.5$) spectrum sources. Thus there is
no evidence for a significant change in source
size between the flat and steep spectrum
population. Interestingly, there are only
four inverted spectrum sources among the flat
spectrum sample, indicating that strongly self-absorbed
 systems are rare among the microjansky
population.

	In Paper I we considered two 
possibilities for the origin of microjansky radio emission in
distant disk galaxies, 1)  increased radio activity
associated with a central engine (e.g., Seyfert
and LINER AGN), and 2) radio emission excited
by star-formation. Both are capable of 
producing flat radio spectra through 
synchrotron self-absorption in the case of the
former, and through increasing amounts of 
thermal radio emission in the later.
These two different physical 
mechanisms take place on very different
physical scales. For the case of star formation
these scales correspond from approximately 0.1-10 kpc
as observed in the local starburst population 
(Condon 1989). Thus the cosmological microjansky
population at a  mean redshift of 0.8
(Paper I) should have an angular
extent of 0.01-1\arcsec ~ if star formation is
the ultimate source of energy powering the
radio emission. On the other hand
if the radio emission has its origin in
an AGN then the flattening of the
spectral index distribution can be attributed to
partial self-absorption. In this case the observed
flux density of the source makes calculation
of a minimum angular size for synchrotron 
self-absorption possible (Pacholczyk 1970). For a 100 $\mu$Jy
source with a critical absorption frequency of
1.4 GHz and an assumed magnetic field strength no 
larger than $10^{-4}$ Gauss, the characteristic
angular size scale is of order $10^{-5}$ arcsec.

	Because the radio spectral index is such a 
sensitive function of absorption varying from -2.5 in 
the case of pure synchrotron self-absorption, to 0.8 for a standard 
transparent, non thermal spectrum, the observation of
low dispersion in the flat spectral indices of the 
microjansky population (i.e., very few inverted spectrum
sources) would suggest that we are seeing very nearly the
same fraction of absorbed radiation in all sources. Perhaps a more
natural explanation for the observed spectral index
distribution of flat spectrum microjansky radio 
sources is that we are observing varying ratios
of thermal to synchrotron emission causing the
spectral indices to vary from -0.1 to 0.8 (cf., Condon
1992). This could be due to the combined effect
of a radio K-correction which serves to bring 
a greater fraction of bremsstrahlung radiation
into the observed radio window for sources
at appreciable redshifts, as well as a
steepening of the
synchrotron radiation itself as a 
result of synchrotron and Compton losses off the
microwave background. A flatenning of the relativistic
electron energy spectrum could flatten the observed
synchrotron spectral index as well.
Sub-arcsecond radio observations of the
microjansky population are necessary to discriminate
between these possibilities (Muxlow et al. 1998).
High frequency observations
would also be useful to determine the slope
of the radio spectrum in these distant radio
sources, but will not be feasible until the 
commisioning of the Millimeter Array or
until the expansion of the high frequency 
capabilities of the VLA.


\section{Conclusions}

	We have presented a complete catalog of 371
radio sources brighter than 40 $\mu$Jy
at 1.4 GHz in a 0.3 deg$^2$ field centered on
the Hubble Deep Field. This is the most
sensitive survey available 
 at this resolution (2\arcsec ) and frequency.
For a subsample of these sources we have 
calculated two point spectral indices based
on 8.5 GHz mosaic observations.

	 The principal
results of this study are:

1. We have extended the direct source count
   at 1.4 GHz to 40 $\mu$Jy, confirming
   that the differential slope for radio
   sources remains steep at $\gamma $ = -2.4
   to this level.

2. The average angular size for the microjansky
   population is observed to decrease as
   a function of flux density. The mean
   size for radio sources between 40-1000 $\mu$Jy
   is 2.0\arcsec , consistent with their association with
   large disk galaxies at $z \sim $ 1.

3. Microjansky radio sources appear to be
   clustered on scales of 1\arcmin - 40\arcmin ,
   corresponding to projected distances of
   0.5 - 20 Mpc .
   
4. The average spectral index for a 1.4 GHz
   selected subsample is $\overline{\alpha } _{1.4} = 0.85
    \pm 0.2$, indicating optically
   thin synchrotron emission as the 
   dominant radio emission mechanism.
   For a 8.5 GHz selected sample, the 
   mean is $\overline{\alpha } _{8.5} = 0.4 \pm 0.1.$
   This flattening of the spectral index
   distribution over that of samples 
   selected above a few millijansky
is consistent with
   either the cosmological evolution of
   the disk galaxy AGN population (LINERs
   and Seyferts) or of their star-formation
   activity. We suggest that we are
   observing increasing amounts of 
   bremsstrahlung radiation in these
   sources, causing the observed decrease
   in the spectral index distribution
   at 8.5 GHz.
 
\begin{acknowledgements}

	I thank my collaborators Ed Fomalont, Ken Kellermann,
Bruce Partridge, Rogier Windhorst, and Tom Muxlow for
their help with this project. This study would not
have been possible without the expert assistance
of the NRAO staff in the planning,
execution, and analysis of these observations, especially
F. Owen and M. Rupen. This
work benefited from suggestions by
J. Condon and J. Wall. I also thank L. Cowie and A. Barger
for pointing out a positional inconsistency
in an earlier version of this work.

\end{acknowledgements} 

\newpage

\section{References}

\nref Cillegi, P. et al. 1998, MNRAS, in press

\nref Clark, B. G. 1980, A\& A, 89, 377

\nref Coleman,P. H. \& Condon, J. J. 1985, AJ, 90, 1431

\nref Condon, J. J. 1989, ApJ, 338, 13

\nref Condon, J. J. 1992, ARA \& A, 30, 575

\nref Condon, J. J. 1997, PASP, 109, 166

\nref Cornwell, T. J. \& Perley, R. A. 1992, A \& A, 261, 353  

\nref Cress, C., Helfand, D., Becker, R., Gregg, M. \& White, R.
  1996, ApJ, 473, 7

\nref Donnelly, R. H., Partridge, R. B \& Windhorst, R. A. 1987, ApJ, 321, 94

\nref Feigelson, E. D. \& Nelson, P. I. 1985, ApJ, 293, 192

\nref Fomalont, E. B., Windhorst, R. A., Kristian, J. A. \& Kellermann,
K. I.
      1991, AJ, 102, 1258

\nref Fomalont, E. B., Goss, W. M., Lyne, A. G., Manchester, R. N. \&
   Justtanont, K. 1992, MNRAS, 258, 497

\nref Fomalont, E. B., Kellermann, K. I., Richards, E. A., Windhorst, R.
A. \&
  Partridge, R. B. 1997, ApJL, 475, 5

\nref Gruppioni, C., Zamorani, G., De Ruitter, H. R., Parma, P, Mignoli, M. \& Hasinger, G
1997, MNRAS, 286, 470

\nref Hammer, F., Crampton, D., Lilly, S. J., LeFevre, O. \& Kenet, T.
1995,
  MNRAS, 276, 1085

\nref Hopkins, A., Afonso, J., Cram, L. \& Mobasher, B. 1999, ApJL,
   in press

\nref Hopkins, A. M., Mobasher, B., Cram, L. \& Rowan-Robinson, M.
  1998, MNRAS, 296, 839

\nref Isobe, T. \& Feigelson, E. D. 1991, ASURV, Rev. 1.2

\nref Katgert, P., Oort, J. \& Windhorst, R. 1988, A\& A, 195, 21

\nref Ling, E. N., Frenk, C. S. \& Barrow, J. D. 1986, MNRAS, 223, 21p

\nref Magliocchetti, M., Maddox, S., Lahav, O. \& Wall, J. 1998,
  MNRAS, 300, 257

\nref Mitchell, K. J.  \& Condon, J. J. 1985, AJ, 90, 1987

\nref Muxlow et al. 1999, {\em in preparation}

\nref Oort, J. A. 1987, A \& AS, 71, 2210

\nref Oort, J. A. 1988, A \& A, 193, 50

\nref Oort, J. A. \& Windhorst, R. A. 1985, A\&A, 145, 4050

\nref Pacholczyk 1970, Radio Astrophysics (San Francisco: W. H. Freeman)

\nref Peebles, P. J. E. 1980, The Large Scale Structure of the Universe 
  (Princeton: Princeton University Press)

\nref Richards, E. A. 1996, in IAU 175: Extragalactic Radio Sources, eds.,
      Ekers, R., Fanti, C. \& Padrielli, L., 593

\nref Richards, E. A., Kellermann, K. I., Fomalont, E. B.,
      Windhorst, R. A., \& Partridge, R. B. 1998a, AJ, 116, 1039 (Paper I)

\nref Richards, E. A. 1999b, et al., {\em in preparation} (Paper III)

\nref Rowan-Robinson, M. {\em et al.} 1997, MNRAS, 289, 490

\nref Windhorst, R. A, Fomalont, E. B., Kellermann, K. I., Partridge, R.
B.,
      Richards, E. A., Franklin, B. E., Pascarelle, S. M.
       \& Griffiths, R. E. 1995,
      Nature, 375, 471

\nref Windhorst, R. A., Fomalont, E. B., Partridge, R. B. \& Lowenthal,
J. D.
      1993, ApJ, 405, 498

\nref Windhorst, R. A., van Heerde, G. M. \& Katgert, P. 1984, A \& AS,
58, 1

\newpage

\section{Figure Captions}

1. A greyscale of the inner 9.5\arcmin
   $\times$ 9.5\arcmin ~
   of the 1.4 GHz image.
   The pixel range is from -10 - 20 $\mu$Jy.
   The effective resolution is 1.8\arcsec ~
   with an rms noise of 7.5 $\mu$Jy. The image has been
   corrected by the primary beam attenuation.

2. Here we show the same 9.5\arcmin $\times$ 9.5\arcmin
   area ~at 8.5 GHz. The greyscale extends from
   -2 - 6 $\mu$Jy. The effective resolution is 3.5\arcsec ~
   with an rms noise of 1.6 $\mu$Jy. This image
   has been corrected by the primary beam attenuation
   and hence the noise is not uniform.

3. The total off-axis response of a point source due to
   the combined effects of bandwidth smearing, time
   averaging smearing, and the primary beam correction
   (gain). The relative amplitude decrease due
    to time average smearing (tsmear), bandwidth
    smearing (bwsmear), and the primary beam attenuation
    (pbcor) are also shown.

4. The angular size distribution for the 1.4 GHz complete
   sample. Notice the increasing number of upper limits
   at lower flux density levels.

5. The 1.4 GHz median angular size vs. flux density
   relation for microjansky radio sources.
   Points from Oort (1988; O88), Coleman
   \& Condon (1985; CC), and this study are
   plotted. The
   curve shown has been calculated according to the
   model given in \S 4 . The broken line represents
   a median angular size independent of flux density.

6a. The 1.4 GHz source counts from this field and other
   deep radio surveys are shown. The count is presented
   in differential form normalized to the counts
   expected in a Euclidean geometry.
   For comparison, counts from Mitchell \& Condon
   (1985; MC), the Phoenix Deep Field (Hopkins et al.
   1998; PDF), Oort \& Windhorst (1985, OW85), and
   the ELAIS survey (Cillegi et al. 1998; ELAIS)
   are plotted. The best fit to the earlier 
   deep survey compilation by Katgert et al. (1988)
   is shown as a solid line.

6b. A blow-up of the sub-millijansky counts. 
   The solid line is the best fit to
   the current survey data. There is good evidence
   for field to field fluctuations above that
   of the statistical noise.

7. The correlation function of the HDF radio 
   sruvey is shown as heavy dots. The correlation
   function in the Lynx3 survey of Oort (1987) 
   is shown for comparison.

8. The spectral index flux density distribution for
   the 1.4 GHz selected sample. The mean of
    the sample as calculated by survival analysis
   is shown as the broken line.

9. The spectral index flux density distribution for
   the 8.5 GHz selected sample. The mean of
    the sample as calculated by survival analysis
   is shown as the broken line.

\bn

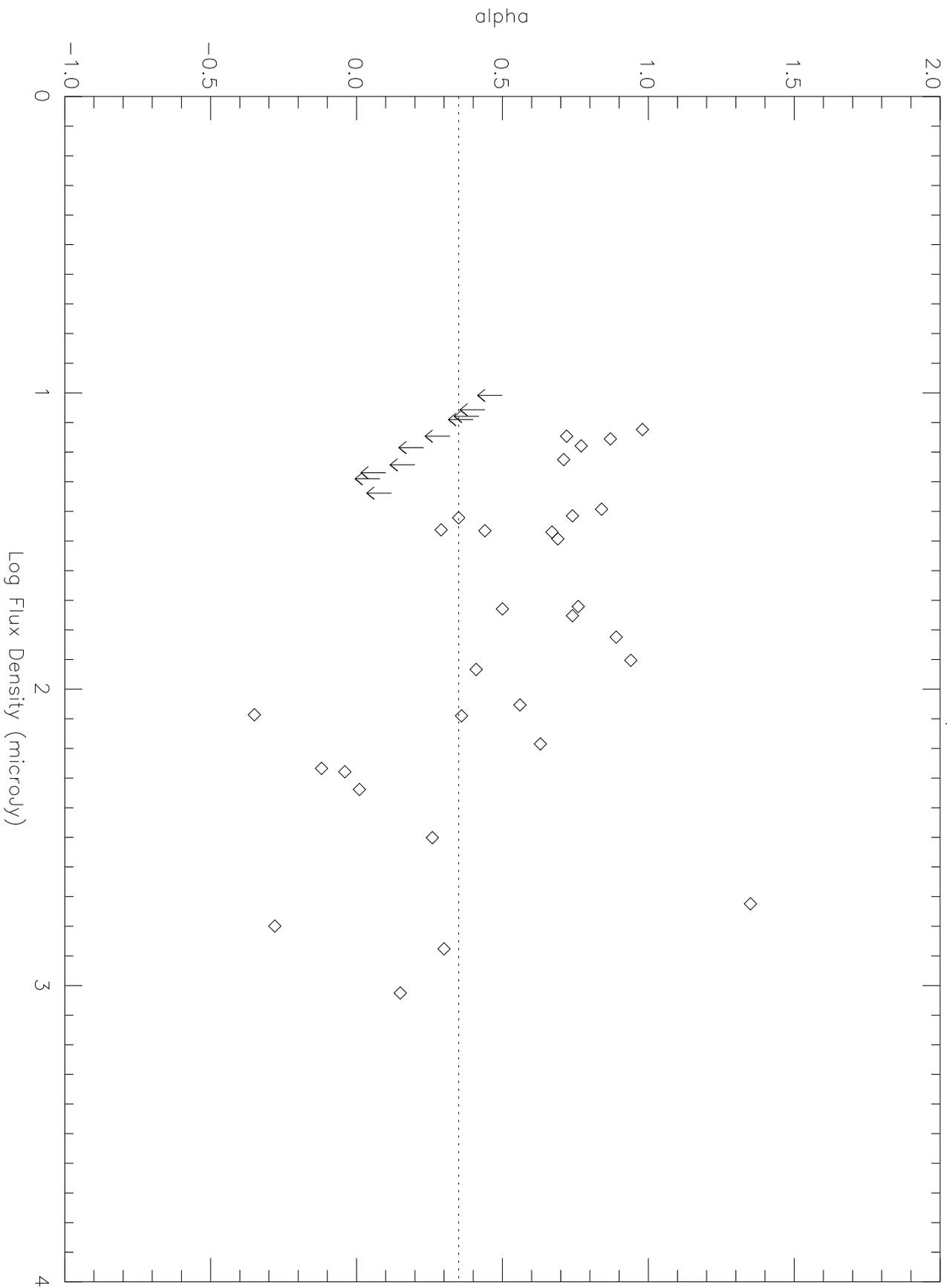
\end{document}